\documentclass{article}

\usepackage{spconf,graphicx}
\usepackage{hyperref}
\usepackage{amsfonts,amsmath,amssymb}
\usepackage{array,booktabs,multirow}
\usepackage{subcaption}     % create sub-tables and sub-figures
\usepackage{bold-extra}     % enable \texttt{\textbf{}}
\usepackage{bm}             % enable bold font in math mode
\title{R{\small escore}BERT: Discriminative Speech Recognition Rescoring with BERT}

\name{\begin{tabular}{c} Liyan Xu$^{1,2}$ \qquad Yile Gu$^{1}$ \qquad Jari Kolehmainen$^{1}$
\qquad Haidar Khan$^{1}$ \qquad Ankur Gandhe$^{1}$ \\
Ariya Rastrow $^{1}$ \qquad Andreas Stolcke$^{1}$ \qquad Ivan Bulyko$^{1}$%
\thanks{$^2$Work done as an applied scientist intern at Amazon Alexa.}
\end{tabular}}

\address{$^1$Amazon Alexa AI, USA \quad  $^2$Emory University, USA}

\begin{document}

\ninept
\sloppy

\maketitle

\begin{abstract}
Second-pass rescoring is an important component in automatic speech recognition (ASR) systems that is used to improve the outputs from a first-pass decoder by implementing a lattice rescoring or $n$-best re-ranking. While pretraining with a masked language model (MLM) objective has received great success in various natural language understanding (NLU) tasks, it has not gained traction as a rescoring model for ASR. Specifically, training a bidirectional model like BERT on a discriminative objective such as minimum WER (MWER) has not been explored. Here we show how to train a BERT-based rescoring model with MWER loss, to incorporate the improvements of a discriminative loss into fine-tuning of deep bidirectional pretrained models for ASR. Specifically, we propose a fusion strategy that incorporates the MLM into the discriminative training process to effectively distill knowledge from a pretrained model. We further propose an alternative discriminative loss. This approach, which we call RescoreBERT, reduces WER by 6.6\%/3.4\% relative on the LibriSpeech clean/other test sets over a BERT baseline without discriminative objective. We also evaluate our method on an internal dataset from a conversational agent and find that it reduces both latency and WER (by 3 to 8\% relative) over an LSTM rescoring model.

\end{abstract}


\section{Introduction}
\label{sec:intro}

The two-pass paradigm has been widely adopted in state-of-the-art ASR systems \cite{deliberation-nips,two-pass,deliberation2020,ankur,hu2021transformer}, where the first pass generates n-best hypotheses, and the second pass reranks them. For the second-pass rescoring models, discriminative training with MWER (minimum WER) objective is typically applied \cite{deliberation2020,mwer2016,mwer2018,ankur} to improve performance, such that the model learns to prefer hypotheses with the lowest WER. %MWER can be either completely separated from LM scoring or jointly trained.

Previous work with discriminative training uses causal language models (CLMs), such as LSTMs or Transformer LMs. While pretrained masked language models (MLMs) such as BERT \cite{bert} have been highly successful on various natural language understanding (NLU) tasks, they have not been widely applied in second-pass ASR rescoring.
Meanwhile, recent studies have shown promising results using BERT in several rescoring studies \cite{mlm-scoring1,mlm-scoring2,chiu2021innovative}, as BERT is pretrained with large corpora and encodes the full hypothesis context using a deep bidirectional model architecture. In particular, previous work~\cite{mlm-scoring1} shows that deep bidirectional Transformers, such as BERT, can outperform their unidirectional counterparts (either forward text, backward text, or the two models combined). Another paper~\cite{mlm-scoring2} shows that a pretrained BERT model that is then fine-tuned on LibriSpeech data can outperform BERT trained from scratch on LibriSpeech clean/other test sets by 4.4\%/3.2\% WER relative, demonstrating the effectiveness of pretraining in BERT. They also show that BERT can outperform GPT \cite{GPT} with comparable model size and pretraining data, which the authors argue is due to the bidirectional nature of BERT.

\begin{figure}[t]
\centering
\includegraphics[width=7cm]{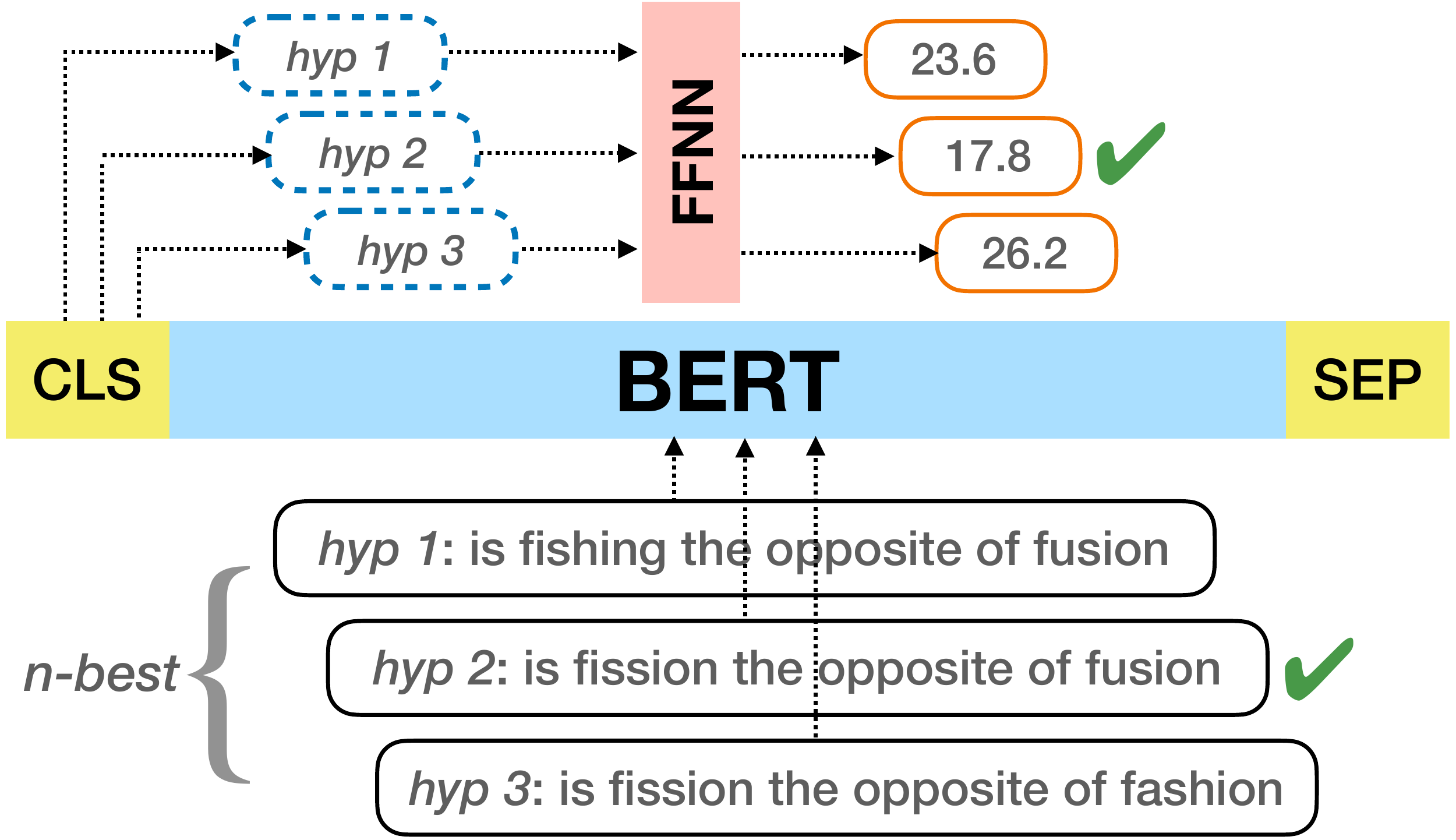}
\caption{Illustration of discriminative scoring with BERT on n-best hypotheses. Each hypothesis is individually encoded by BERT and represented by CLS; it is then followed by a feed-forward NN to compute a sentence-level second-pass LM score. The scores are then interpolated with first pass scores for reranking.}
\label{fig:bert}
\vspace{-2ex}
\end{figure}

In this work, we propose a method to train BERT-style rescoring models with a discriminative objective, to leverage the aforementioned benefits from both approaches. Typically, pseudo log-likelihood (PLL)~\cite{mlm-scoring1,mlm-scoring2}---the sum of the negative log-likelihoods of each individual token given the bidirectional context---is used to rescore n-best output to improve WER, a computationally expensive process, particularly for longer sentences. For discriminative training, this issue is exacerbated as the PLL computation needs to be repeated for each hypothesis individually. The previous work~\cite{mlm-scoring2} solves this issue by distilling the PLL into a single score prediction at the start-of-sentence token (CLS).
In this work, illustrated in Figure~\ref{fig:bert}, we extend this approach and use the score from the CLS representation to perform discriminative training, as discussed in Section~\ref{sssec:dis_only}, with either MWER loss or a novel discriminative training loss dubbed matching word error distribution (MWED), described in Section~\ref{subsec:loss_function}. Finally, in Section~\ref{sssec:mlm_dis} we propose a fusion strategy that incorporates the MLM into the discriminative training process, giving further improvements.

%Specifically, we introduce our first model in Section~\ref{subsec:bert-mwer} that adapts BERT with the MWER objective in a simple architecture, utilizing the deep bidirectional encoding power of BERT in the supervised MWER training.
%We further explicitly fuse the MLM perspective into the MWER process by applying a fusion strategy described in Section~\ref{subsec:fusing-mlm}.
%Upon that, we also propose a novel MWER training loss dubbed as Matching Word Error Distribution (MWED) in Section~\ref{subsec:mwed} that outperforms the existing loss on certain datasets.
%
We name the aforementioned approach RescoreBERT, and evaluate it on four datasets covering multiple domains and locales. Results show that discriminative training significantly improves upon non-discriminative BERT rescoring, on all test sets. The new MWED training loss is found to be a strong alternative to MWER. The results also show that the fusion approach for incorporating MLM into discriminative training can further improve WER. Lastly, to achieve lower latency for streaming applications, we develop a method to further distill the model while maintaining WER improvements.

%On LibriSpeech public datasets, it has already accomplishes 5-11\% relative improvement over the baseline on three internal datasets, and obtains comparable results on two public LibriSpeech \cite{libri} test sets due to its limited MWER training data. Combining MLM perspective into MWER further robustly improves performance on all datasets, especially outperforming the baseline on two LibriSpeech test sets by relatively 5.3\% and 1.7\%. The new MWED loss is shown to yield the best performance for LibriSpeech (6.6\% and 3.4\%) and one internal dataset in English (6.6\%), while the previous MWER loss performs the best on two other datasets in Japanese (12.0\% and 12.7\%).
%Lastly, we also conduct a latency study in Section~\ref{subsec:latency} that employs much smaller distilled BERT, demonstrating that smaller BERT can maintain $\sim$75\% improvement while having a production-ready latency.

% \input{tex/related}
\section{Approach}
\label{sec:approach}

\subsection{BERT Without Discriminative Training}
\label{subsec:without_mwer}

In this section, we review previous work on BERT rescoring models that are not trained with discriminative objective.

\subsubsection{Pseudo log-likelihood (PLL)}
\label{sssec:pll}

Let $E = (e_1, ..., e_{|E|})$ be a sequence of tokens. Similar to the log-likelihood of a sequence as is commonly used in CLM scoring \cite{shallow-deep-fusion,fusion,clm-scoring}, the pseudo log-likelihood (PLL) of a sequence using an MLM, first introduced by \cite{mlm-scoring1}, is defined as:
\begin{align}
    \text{PLL}(E) = -\sum_{t=1}^{|E|} \log P(e_t | E_{\setminus t}) \label{eq:pll}
\end{align}
where $E_{\setminus t} = (..., e_{t-1}, \text{[MASK]}, e_{t+1}, ...)$ is the sequence whose corresponding position is replaced by the [MASK] token used in MLM pretraining. The PLL is thus the sum of the negative log-likelihoods of each token given the bidirectional context, with lower scores indicating more probable sequences.

\subsubsection{MLM distillation (MD)}
\label{sssec:md}

Although the PLL demonstrates good performance for second-pass rescoring \cite{mlm-scoring1}, it is not computationally efficient: $|E|$ additional sequences masking every position need to be generated and encoded by BERT, thus the computation due to PLL is on the order of $|E|$ times that of an autoregressive Transformer model of similar size.
Following \cite{mlm-scoring2}, one can ``distill'' the PLL calculation into a single utterance-level score using the CLS representation, such that the model is able to approximate PLL, while eliminating the need for masking $|E|$ times, as well as the large vocabulary softmax in $P(e_t | E_{\setminus t})$, thereby reducing computation significantly.
As shown in the equations below, each sentence $E_i$ is  individually encoded by BERT, represented by the hidden state of the CLS token in the last Transformers layer, denoted by $g_i$. An additional layer is stacked on top of the CLS hidden states to produce the score $s^l_i$ for $E_i$. The distillation is achieved by training the model to mimic PLL scores using mean squared error (MSE) regression loss:
\begin{align}
 g_i &= \text{BERT}^{\text{CLS}}(E_i) \label{eq:cls} \\
    s^l_i &= \text{FFNN}(g_i)  \label{eq:cls_score}\\
    \mathcal{L}_{\text{MD}} &= |s^l_i - \text{PLL}(E_i)|^2
\end{align}
FFNN denotes the learnable feed-forward neural network, $s^l_i$ is the predicted PLL approximation, and PLL$(E_i)$ is precomputed offline using Eq.~\eqref{eq:pll}. Note that the PLL can be computed by a larger \emph{teacher} model.

\subsection{BERT With Discriminative Training}
\label{subsec:with_mwer}

%In this section, we propose three methods to train BERT with discriminative objective: \emph{MWER Only}, \emph{MD-MWER}, and \emph{MD-MWED}.
We now propose methods for training BERT with discriminative objective functions.
%We will first discuss discriminative loss functions, and then methods to train models with only discriminative loss or combined discriminative and MLM loss.
% (1) MWER: From a pretrained BERT model, we simply finetune itl so that score from CLS presentation minimizes MWER loss. (2) MD-MWER: From a pretrained BERT model, we first finetune using MD, before finetuning again with combined MD and MWER loss. (3) MD-MWED: Same as MD-MWER, except that a new loss function MWED is used.
For any utterance, let $\vec{E} = \lbrace E_1, ..., E_n \rbrace$ be the n-best hypotheses obtained from beam search in the first-pass decoder. For any $E_i \in \vec{E}$, let $s^a_i$ be its given score from the first pass, and $s^l_i$ be the score from the second pass (same as Eq.~\eqref{eq:cls_score}), with lower scores for more likely hypotheses for both; let $\epsilon_i$ be its number of word errors (edit distance) from the ground truth transcription.
%Each hypothesis $E_i$ is then individually encoded by BERT, represented by the hidden state of the CLS token in the last Transformers layer, denoted as $g_i$. An additional layer is stacked on top of CLS hidden states to produce the score $s^m_i$ for $E_i$. 
Following the common theme of second pass rescoring approaches, the final score $s_i$ is the linear combination of the first-pass and second-pass scores:
\begin{align}
%    s^m_i &= \text{FFNN} (g_i, E_i) \label{eq:sm}\\
    s_i &= s^a_i + \beta \cdot s^l_i \label{eq:bert-mwer} \quad,
\end{align}
where $\beta$ is the hyperparameter controlling the second-pass contribution. $s_i$ is then used to compute discriminative loss, as defined next.

\subsubsection{Discriminative loss function}
\label{subsec:loss_function}
We explore two discriminative loss functions: MWER and MWED.

\textbf{MWER (Minimum word error rate)}:
%\paragraph{MWER}
A standard discriminative loss function for ASR rescoring is MWER~\cite{mwer2018}. The training minimizes the expected number of word errors for the n-best hypotheses:
\begin{align}
    P_i &= \frac{e^{-s_i}}{\sum_{j=1}^n e^{-s_j}} \\
    \overline{\epsilon}_H &= \frac{1}{n} \sum_{i=1}^n \epsilon_i \\
    \mathcal{L}_{\text{MWER}} &= \sum_{i=1}^n P_i \cdot (\epsilon_i - \overline{\epsilon}_H) \label{eq:mwer-loss}.
\end{align}
$P_i$ is the posterior probability of each hypothesis, normalized over the hypotheses list from the first pass, such that higher probabilities indicate perferred hypotheses. $s_i$ is the final score of the hypothesis as in Eq.~\eqref{eq:bert-mwer}. The MWER loss $\mathcal{L}_{\text{MWER}}$ represents the expected number of relative word errors, with $\overline{\epsilon}_H$ being the averaged word errors across the n-best list, which does not change the optima but helps to reduce the variance. 

\textbf{MWED (Matching word error distribution)}:
MWED is a new loss function proposed here. Its goal is to mimic the distribution of n-best word errors through the predicted scores. As a result, the ranking of final scores should ideally be exactly the same as ranking by the word errors, which could potentially lead to better score interpolation at evaluation. By contrast, the model trained with the existing MWER loss as in Eq~\eqref{eq:mwer-loss} picks the best hypothesis discriminatively, such that the full probability mass should be assigned to the one with minimum word errors in the ideal case. 

%MWED is a new loss function proposed here. Its goal is to correlate the distribution of n-best word errors with the predicted scores, instead of rewarding correctly picking the single best hypothesis.
%As a result, the ranking of the learned scores should ideally be the same as the ranking by word errors, which could potentially lead to better score interpolation at evaluation. By contrast, the model trained with the existing MWER loss as in Eq~\eqref{eq:mwer-loss} picks only the best hypothesis, such that the full probability mass would be assigned to the minimum-word-error hypothesis and no discrimination happening among lower-ranked ones.

%We first transform $s_i$ so that the total mass of predicted scores among n-best is equal to that of word errors, achieved by the softmax temperature $T = \sum_{i=1}^n s_i / \sum_{i=1}^n \epsilon_i$. 
The MWED loss is proposed as the following:
\begin{align}
    d^\epsilon_i &= \frac{e^{\epsilon_i}}{\sum_{j=1}^n e^{\epsilon_j}} \\
    d^s_i &= \frac{e^{s_i / T}}{\sum_{j=1}^n e^{s_j / T}} \\
    \mathcal{L}_{\text{MWED}} &= -\sum_{i=1}^n d^\epsilon_i \log d^s_i \label{eq:mwed}
\end{align}
$d^\epsilon_i$ and $d^s_i$ represent the relative distribution of word errors and predicted scores over the n-best list. $\mathcal{L}_{\text{MWED}}$ is the cross-entropy from scores to word errors, equivalent to optimizing the Kullback–Leibler divergence between the two distributions. Due to that $s_i$ contains $s^a_i$ which is fixed, to stabalize the match of the two distributions, we add a hyperparameter $T$ to rescale the distribution mass of $s_i$. In practice, we found that $T = \sum_{i=1}^n s_i / \sum_{i=1}^n \epsilon_i$ can yield good performance. 

\subsubsection{Training with discriminative loss only}
\label{sssec:dis_only}
Training BERT naively with discriminative loss using word-level scores, as done in~\cite{ankur, mwer2016}, requires computation of Eq.~\eqref{eq:pll} for every hypothesis and is prohibitively expensive during both training and inference.
Instead, it can be fine-tuned such that the sentence-level score from the CLS representation (as in Eq.~\eqref{eq:cls_score}) minimizes the discriminative loss $\mathcal{L}_{\text{MWER}}$ or $\mathcal{L}_{\text{MWED}}$ defined earlier. 
 
In Section \ref{sec:results}, we show results using this approach with $\mathcal{L}_{\text{MWER}}$, labeled ``\emph{MWER Only}'', where we perform MWER training on a pretrained BERT with domain adaptation.
%Each hypothesis in this work is encoded and predicted independently without awareness of other candidates.
%Further interactions could also be added to strengthen the representation, such as applying attentions among n-best, which we will leave for future work.
% We employ one (or more) self-attention layer(s) \cite{attention} to perform contextualized encoding on the CLS representation of n-best hypotheses:
% \begin{align}
%     g'_i &= \text{SELF-ATTN}(g_i | g_1, \dots, g_n) \label{eq:sa}
% \end{align}
% $g'_i$ will simply replace $g_i$ in Eq~\eqref{eq:sm}. $\phi(E_i)$ can further include additional meta features such as the first-pass ranking of $E_i$ in its n-best, since each hypothesis now has access to other candidates.

%The above model that adapts BERT with MWER can already serve as a strong rescoring model, but no explicit information of LM has yet been exploited in the process. While previous work has taken CLM models and jointly trained both MWER and CLM objectives \cite{deliberation2020,ankur,mwer2018}, BERT is pretrained with MLM that cannot be directly adopted in the MWER process. We fuse MLM into MWER through the following strategy described below.
%
\subsubsection{Training with combined MLM and discriminative loss}
\label{sssec:mlm_dis}
We propose a fusion strategy to incorporate MLM distillation into discriminative training. It is accomplished by making two modifications to the approach in Section~\ref{sssec:dis_only}, where only discriminative loss is applied.

First, we apply a pretraining step using MD alone on a large-scale text-only corpus, so that the discriminative training can be warm-started from a better initialization point. Unlike MWER training, MD only needs text-only data and their PLL scores computed by a \emph{teacher} model. Therefore, the distillation itself can be trained on much more data than the n-best hypotheses used in MWER training. 

Second, we introduce the new loss $\mathcal{L}$ to replace $\mathcal{L}_{\text{MWER}}$ or $\mathcal{L}_{\text{MWED}}$ in the discriminative training step:
% We further incorporate MD into MWER training and add $s^l_i$ as part of the second-pass score:
\begin{align}
   % s_i &= s^a_i + \beta \cdot \big( s^l_i + s^m_i \big) \label{eq:final-score} \\
    \mathcal{L} &= \mathcal{L}_{\text{Discriminative}} + \lambda \cdot \ \sum_{i=1}^n \mathcal{L}_{\text{MD}} (E_i)\label{eq:joint},
\end{align}
where $ \mathcal{L}_{\text{Discriminative}}$ is the discriminative loss that can be either  $\mathcal{L}_{\text{MWER}}$ or  $\mathcal{L}_{\text{MWED}}$. $\mathcal{L}_{\text{MD}}$  is similar to cross-entropy regularization added to MWER loss in \cite{mwer2016}, and controlled by the hyperparameter $\lambda$.
It is found that MD pretraining is more important than adding additional MD loss. On top of MD pretraining, having an additional step of adding MD loss yields less than 0.5\% relative improvement from all experiments.
%Eq~\eqref{eq:final-score} now explicitly predicts the LM score $s^l_i$, and the MWER score $s^m_i$ that further fills the gap to minimize word errors.
%Note that we could also simply dismiss $s^m_i$ with a small $\lambda$, and we do not observe performance degradation. For clarity, we do include $s^m_i$ in our experiments for better interpretability.

In Section \ref{sec:results}, we show the results using this approach with $\mathcal{L}_{\text{MWER}}$ and $\mathcal{L}_{\text{MWED}}$, which are named \emph{MD-MWER} and \emph{MD-MWED}, respectively.

\section{Experiments}
\label{sec:experiments}

\subsection{Datasets}
\label{subsec:datasets}

We evaluated our approach on four datasets in multiple domains and locales to test its general applicability, including one public dataset LibriSpeech \cite{libri} and three internal dataset based on a converstional agent (one for Information (Info) domain in English (en), and two for Info and Navigation (Nav) domains in Japanese (ja)).

For LibriSpeech, an LAS model \cite{las} is adopted as the first-pass decoder, and we use the same decoded 100-best hypotheses of the dev and test set along with their first-pass scores used by \cite{mlm-scoring1,mlm-scoring2}. Since there is no dedicated training set provided for MWER, we combine the decoded hypotheses from both dev-clean and dev-other as the MWER training set, and randomly hold out 10\% utterances as the MWER dev set. The resulting training/dev set has 5011/556 utterances, with up to 100 hypotheses per utterance. For MLM distillation, we sample 4 million utterances from the in-domain text corpus provided by LibriSpeech as the training set, similar to \cite{mlm-scoring2}.

All internal datasets consist of de-identifed live user interactions with a conversational agent, decoded by a RNN-T model \cite{rnnt} for en or a hybrid HMM model for ja. For Info (en), $\sim$100/5 hours of utterances are used as the MWER training/dev set; the test set has $\sim$2 hours of long-tail info utterances.
% The n-best size (number of hypotheses per utterance) is 6.6 on average.
For Info and Nav in ja, we use a single MWER training/dev set that consists of $\sim$220/10 hours of utterances in multiple domains including Info and Nav domains. The test set has $\sim$1/3 hours of utterances for Info/Nav respectively.
% and the averaged n-best size is 4.9. 
For MD, 4 million in-domain utterances sampled from user interactions are used as training set.
% For MLM distillation, we sample 4 million in-domain utterances from user interactions that include gold transcriptions as well as filtered one-best hypotheses (non-gold) from the first-pass.

\subsection{Implementation}
\label{subsec:implementation}

For LibriSpeech, we use the uncased BERT\textsubscript{Base} for our experiments to enable direct comparison with previous work. For internal datasets, we use an in-house multilingual BERT of $\sim$170M parameters (excluding embedding size) with 16 layers and 1024 hidden size, supporting both en and ja locales. In addition, our final MWER setting also includes two smaller BERT models of $\sim$17M/5M parameters distilled~\cite{distillation} from the 170M BERT model, both with only 4 layers and 768/320 hidden size respectively. All BERT models are implemented in PyTorch and pretrained on public data. We limit the maximum sequence length to 128 for LibriSpeech, and 32 for others; longer utterances will be truncated. The $\lambda$ parameter in Eq.~\eqref{eq:joint} is set to $10^{-4}$. We found that $\lambda$ between $10^{-4}$ and $10^{-3}$ generally yield same performance.

Before we conducted any training, we first performed domain adaptation for BERT, as BERT contains general world knowledge from pretraining but not necessarily task-specific knowledge. Therefore, we take the pretrained BERT and further train with the MLM objective on the in-domain corpus. For LibriSpeech, we train 400K steps on the provided text corpus, similar to \cite{mlm-scoring2}. For internal datasets, we train 200K steps on the in-domain transcriptions for each of the en and ja locales. Each step has an effective batch size of 256 utterances for both of these cases.

\subsection{Baseline and Evaluation Protocol}
\label{subsec:baseline}

For LibriSpeech, we use the results of the MLM distillation (MD) as the baseline, which can be seen as our re-implementation of the ``sentence-level fine-tuning'' results from \cite{mlm-scoring2}, which has the same low-latency scoring as our MWER setting. We also provide the results of high-latency PLL scores for comparison.
WER is used as the evaluation metric, and the optimal interpolation weight $\beta$ in Eq.~\eqref{eq:bert-mwer} is linearly searched on the dev set.

For internal datasets, we use the LM scoring from a 2-layer LSTM trained with noise contrastive estimation (NCE) \cite{alexa-lstm} as the baseline, which is often employed in industrial settings for streaming applications. The LSTM is trained on the same data used for domain adaptation for BERT.
New scores from BERT replace the existing LSTM scores, and the optimal weight is searched on the dev set as well. We report the relative improvements in WER for en, and in CER (character error rate) for ja.

\section{Results and Analysis}
\label{sec:results}

\subsection{Comparing Different BERT Rescoring Approches}

Tables~\ref{tab:results}(a) and (b) show the evaluation results on both LibriSpeech and internal datasets. Here, \emph{PLL} denotes the approach in Section~\ref{sssec:pll} which is computationally expensive; \emph{MD} denotes the approach in Section~\ref{sssec:md} that distills the PLL score; \emph{MWER Only} denotes the approach described in Section~\ref{sssec:dis_only} that trains BERT with the MWER objective only; \emph{MD-MWER} and \emph{MD-MWED} denote the approaches described in Section~\ref{sssec:mlm_dis} that incorporate MLM into discriminative training with MWER and MWED loss functions (as in Section~\ref{subsec:loss_function}), respectively.
% In practice, we observe that the performance boost from the n-best contextualization as in Eq~\eqref{eq:sa} is not consistent possibly due to overfitting from insufficient MWER training data, therefore, we turn it off in our experiments and leave it as future work when more data is available.
\begin{table}[tbp!]
\caption{Evaluation results on the test partitions of all datasets. Details of the baseline and evaluation protocol are described in Section~\ref{subsec:baseline}.  The case of ``MWED only'' is not included, and the relative difference between it and “MWER only” is within 1\% for all tests.}
\label{tab:results}
\vspace{-1ex}
\begin{subtable}{\columnwidth}\centering
\caption{WER on the two test sets of LibriSpeech using BERT\textsubscript{Base}. Numbers inside parentheses are relative improvements compared to the baseline.}
\label{subtab:results-libri}
\resizebox{0.72\columnwidth}{!}{
\begin{tabular}{l|cc}
\toprule
& Test-Clean & Test-Other \\
\midrule
First-Pass & 7.26 & 20.37 \\
PLL & 4.54 & 16.08 \\
\midrule
Baseline (MD) & 4.67 & 16.15 \\
MWER Only & 4.82 (-3.2\%) & 16.35 (-1.2\%) \\
MD-MWER & 4.42 (5.3\%) & 15.87 (1.7\%) \\
MD-MWED & \bf 4.36 (6.6\%) & \bf 15.60 (3.4\%) \\
\bottomrule
\end{tabular}}
\end{subtable}
\vspace{0.3em}

\begin{subtable}{\columnwidth}\centering
\caption{Relative improvements of WER (for en) and CER (for ja) on three internal datasets, using the in-house 170M BERT model.}
\label{subtab:results-alexa}
\resizebox{0.72\columnwidth}{!}{
\begin{tabular}{l|ccc}
\toprule
& Info (en) & Info (ja) & Nav (ja) \\
\midrule
LSTM & Baseline & Baseline & Baseline \\
MD & 2.6\% & 3.7\% & 5.6\% \\
MWER Only& 5.3\% & 11.8\% & 11.2\% \\
MD-MWER & 4.0\% & \bf 12.0\% & \bf 12.7\% \\
MD-MWED & \bf 6.6\% & 10.4\% & 12.2\% \\
\bottomrule
\end{tabular}}
\end{subtable}
\vspace{-4ex}
\end{table}

Based on these results, we can make three observations. First, discriminative training significantly improves upon non-discriminative BERT rescoring (MD) across all test sets: 6.6\%/3.4\% WER relative improvement on LibriSpeech and 4\%/8.3\%/7.1\% relative WER reduction on internal datasets. What is particularly striking is that on LibriSpeech, because of discriminative training, both MD-MWER and MD-MWED now outperform the much more computationally expensive PLL approach. Second, the fusion approach of incorporating MLM in discriminative training improves on all test sets. The effect is particularly strong on the LibriSpeech test sets, where MWER Only would actually perform worse than MD with 3.2\%/1.2\% relative WER degradations, but where the fusion approach now gives 5.3\%/1.7\% relative improvement over MD. Third, to compare the new loss function MWED with the existing MWER, MD-MWED achieves better performance on both LibriSpeech (1.3\% and 1.7\% relative) and Info (en) (2.6\% relative) over MD-MWER, but worse than MD-MWER on both ja test sets. This result shows that MWED can be a strong alternative loss in the MWER training, and the final performance can be dataset-specific. One potential explanation for MWED being less effective for ja is that the CER distribution is spikier than the WER distribution, resulting in less stable gradients from the relative entropy in Eq.~\eqref{eq:mwed}.

\begin{figure}[b]
\vspace{-2.9ex}
\centering
\includegraphics[width=0.83\columnwidth]{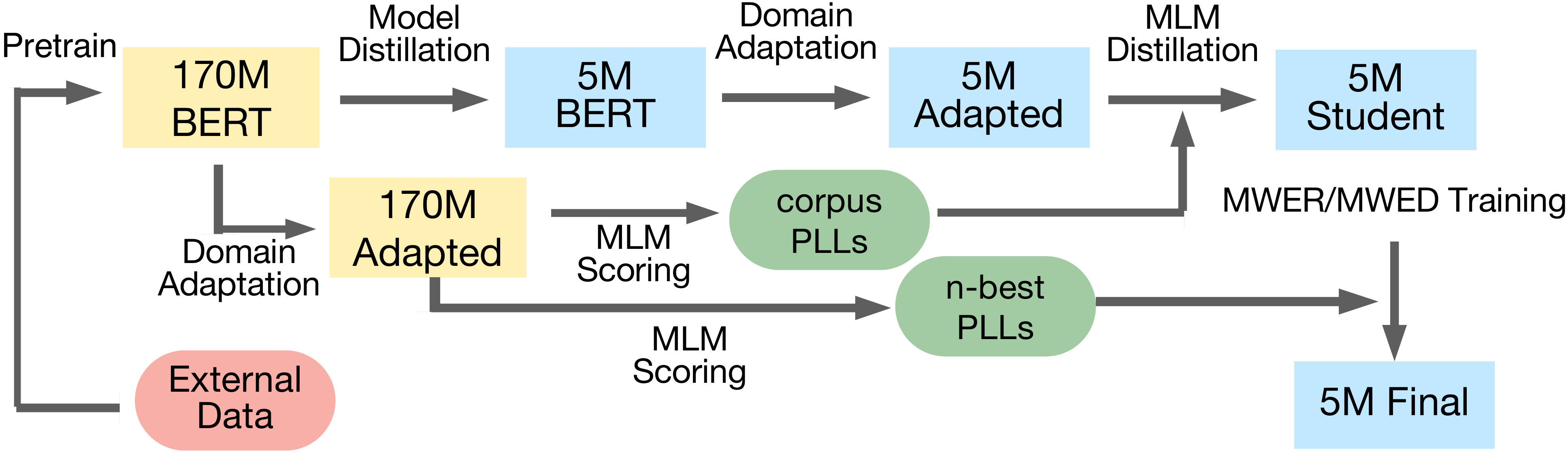}
\caption{Diagram of training the distilled 5M-parameter BERT model with the fusion strategy described in Section~\ref{sssec:mlm_dis}); the 170M BERT model is the \emph{teacher} for calculating PLL scores.}
\label{fig:diagram}
\vspace{-4ex}
\end{figure}

\subsection{Low Latency for Streaming Applications}
\label{subsec:latency}

A main challenge when applying second-pass rescoring for streaming applications is to keep user-perceived latency low while obtaining the accuracy gains. We next examine how to further distill the model to reduce its compute footprint and hence, achieve even lower latency. We focus on the internal datasets for this task. As described in Section~\ref{subsec:implementation}, in addition to the 170M BERT model, we have 17M and 5M BERT models distilled~\cite{distillation} from it. To achieve low latency, we can perform MD-MWER or MD-MWED starting from this smaller model, except that to maintain WER gains, we use the 170M BERT model to compute PLL scores in MD training, as well as in the final combined loss. This training process is illustrated in Figure~\ref{fig:diagram}. % for the example of the 5M BERT model.

Table~\ref{subtab:smaller} shows the relative improvements using BERT models with 170M, 17M, and 5M parameters described in Section~\ref{subsec:implementation} with the best settings for each dataset.
Smaller BERT is expected to yield less improvement; nevertheless, the degradation is within a relatively small margin even for the much smaller 5M version that has only $\sim$3\% the parameters of the 170M model, with still nearly 8\% improvement on two ja datasets and 3+\% on Info (en).
%Future work could also apply model distillation \cite{distillation} in MWER training for further improvement of the smaller BERT models.
% Note that the performance shown in Table~\ref{subtab:smaller} is the lower-bound; we expect more improvement if the model distillation \cite{distillation} is also applied in the MWER training process.

\begin{table}[tbp!]
\caption{Performance and latency study of our approach using BERT of different parameter sizes on three internal datasets.}
\label{tab:latency}
\vspace{-1ex}
\begin{subtable}{\columnwidth}\centering
\caption{Relative improvements over LSTM (4M) on three internal datasets, using the best setting (MD-MWER/MWED) according to Table~\ref{subtab:results-alexa} for each dataset.}
\label{subtab:smaller}
\resizebox{0.6\columnwidth}{!}{
\begin{tabular}{r|ccc}
\toprule
BERT & Info (en) & Info (ja) & Nav (ja) \\
\midrule
170M & \bf 6.6\% & \bf 12.0\% & \bf 12.7\% \\
17M & 3.5\% & 9.1\% & 9.4\% \\
5M & 3.1\% & 7.8\% & 7.8\% \\
\bottomrule
\end{tabular}}
\end{subtable}
\vspace{0.3em}

\begin{subtable}{\columnwidth}\centering
\caption{Averaged latency (in ms) of each batch using 2 threads on a CPU, with a batch size of 5 hypotheses. SL: input sequence/hypothesis length. Parentheses indicate relative latency compared to LSTM.}
\label{subtab:latency}
\resizebox{0.96\columnwidth}{!}{
\begin{tabular}{l|cccc}
\toprule
SL & LSTM (4M) & BERT (5M) & BERT (17M) & BERT (170M) \\
\midrule
16 & 9.7 & \bf 7.6 (78\%) & 17.5 (180\%) & 180 (1.8k\%) \\
32 & 18.7 & \bf 11.0 (59\%) & 26.3 (141\%) & 270 (1.4k\%) \\
\bottomrule
\end{tabular}}
\end{subtable}

\vspace{-4ex}
\end{table}

Table~\ref{subtab:latency} shows the latency comparison among three BERT models and the NCE-based LSTM of 4M parameters, using the PyTorch benchmarking tool under direct model inference in Python using 2 threads on a CPU. 5M BERT is shown to be faster than LSTM, while 17M BERT is slower but also appears comparable. Overall, Table~\ref{tab:latency} shows that our proposed approach can be a superior substitute for LSTM scoring in deployed systems. In particular, from Table \ref{subtab:smaller} and \ref{subtab:latency}, 5M BERT significantly outperforms a similar-sized 4M LSTM, both in WER (by 3.1\%/7.8\%/7.8\%) and in latency (by 22\%/41\%).
%In particular, the 5M BERT model trained with MD-MWER/MWED demonstrates better latency and 3-8\% relative WER reduction over the LSTM.
%\subsection{LM Scoring vs. MWER Scoring}
%\label{subsec:mlm-result}

%Since MD alone can serve for LM scoring \cite{mlm-scoring2}, we compare the evaluation results using the LM scoring versus the MWER training. Table~\ref{tab:mlm} shows the results using BERT with two settings of LM scoring: the high-latency PLL scores, and the low-latency approximated PLL scores from MD. For internal datasets, we also include the smaller 17M BERT. First, PLL mostly outperforms MD as expected; but comparing with MWER in Table~\ref{tab:results}, PLL performs similar to MWER for LibriSpeech, but worse for all internal datasets, with the gap up to 6.8\%. Second, the parameter size may have a larger impact on the final performance for the LM scoring than MWER: the MD process for 17M BERT learns the same PLL scores as 170M BERT, but has 70+\% degradation on Nav (ja), which is not seen in MWER-based models as shown in Table~\ref{subtab:smaller}.

%\input{tab/mlm}

%The above findings suggest that supervised MWER training can be more stable and likely has better performance than the unsupervised LM scoring, however at a cost of human efforts for transcriptions; incorporating the LM perspective (in this case, the proposed fusion strategy) in MWER training can in turn help to further improve the performance, especially when the MWER training data is limited, as shown by the results of LibriSpeech in Table~\ref{subtab:results-libri}.

\section{Conclusion}
\label{sec:conclusion}

We have proposed a method to train a BERT rescoring model with discriminative objective functions. We show that discriminative training can significantly improve BERT rescoring on a variety of datasets:  6.6\%/3.4\% relative WER improvement on LibriSpeech and 4\%/8.3\%/7.1\% relative WER improvement on internal voice assistant datasets. The proposed fusion strategy to incorporate MLM into discriminative training is found to further reduce WER. We also propose a new discriminative loss MWED that is a strong alternative to the standard MWER loss, yielding 1.3\% and 1.7\% relative WER improvement over MWER loss on LibriSpeech and 2.6\% relative improvement on one internal dataset. Lastly, we show how to further distill the model to achieve even lower latency for streaming applications while preserving WER improvements.  We achieve 3-8\% relative WER improvement and lower latency compared to a baseline LSTM model on internal data.
%We propose to use BERT as a deep bidirectional encoder for ASR rescoring. Our approach consists of three phases: the adaptation of BERT in the MWER scoring, the incorporation of MLM into MWER through a fusion strategy, as well as a novel training loss MWED. Evaluation on the public LibriSpeech and three internal datasets show that our approach is able to improve upon the baseline by a large margin;
%with 6.6\%/3.4\% relative improvement for LibriSpeech and 6-12\% for other datasets. Additional analysis shows our approach is 
%further analysis suggests that using much smaller BERT can be a practical solution for production deployment.
% in terms of latency and performance.

% \begin{figure}[htb]
% \begin{minipage}[b]{1.0\linewidth}
%   \centering
%   \centerline{\includegraphics[width=8.5cm]{fig/image1.eps}}
% %  \vspace{2.0cm}
%   \centerline{(a) Result 1}\medskip
% \end{minipage}
% %
% \begin{minipage}[b]{.48\linewidth}
%   \centering
%   \centerline{\includegraphics[width=4.0cm]{fig/image3.eps}}
% %  \vspace{1.5cm}
%   \centerline{(b) Results 3}\medskip
% \end{minipage}
% \hfill
% \begin{minipage}[b]{0.48\linewidth}
%   \centering
%   \centerline{\includegraphics[width=4.0cm]{fig/image4.eps}}
% %  \vspace{1.5cm}
%   \centerline{(c) Result 4}\medskip
% \end{minipage}
% %
% \caption{Example of placing a figure with experimental results.}
% \label{fig:res}
% %
% \end{figure}

% \vfill\pagebreak
\bibliographystyle{IEEEbib}
\bibliography{references}

\end{document}